\UseRawInputEncoding
\documentclass[a4paper,fleqn]{cas-dc}

\usepackage[authoryear,longnamesfirst]{natbib}

\usepackage{graphicx}
\usepackage{longtable}
\usepackage[utf8]{inputenc}

\usepackage{amsmath}
\usepackage{multirow}
\usepackage{xurl} 
\usepackage{hyperref}
\hypersetup{
    breaklinks=true,
    colorlinks=true,
    linkcolor=magenta,
    filecolor=blue,      
    urlcolor=cyan,
    citecolor=green,
}
\usepackage{amsmath,amssymb,amsfonts}
\usepackage{makecell}
\usepackage{textcomp}
\usepackage{calligra}
\usepackage{tabularx}
\usepackage{caption}

\usepackage{adjustbox}
\usepackage{float}
\usepackage{multirow}

\def\tsc#1{\csdef{#1}{\textsc{\lowercase{#1}}\xspace}}
\tsc{WGM}
\tsc{QE}
\tsc{EP}
\tsc{PMS}
\tsc{BEC}
\tsc{DE}

\begin{document}
\let\WriteBookmarks\relax
\def\floatpagepagefraction{1}
\def\textpagefraction{.001}

\shorttitle{AutoLungDx: A Hybrid Deep Learning Approach for Early Lung Cancer Diagnosis Using 3D Res-U-Net, YOLOv5, and Vision Transformers}

\shortauthors{Shuvo et~al.}

\title [mode = title]{AutoLungDx: A Hybrid Deep Learning Approach for Early Lung Cancer Diagnosis Using 3D Res-U-Net, YOLOv5, and Vision Transformers}                      

\author[1]{Samiul Based Shuvo}[orcid=0000-0002-5035-2114,]
\cormark[1]
\ead[URL]{sbshuvo@bme.buet.ac.bd}

\author[1]{Tasnia Binte Mamun}
\ead[URL]{tasnia@bme.buet.ac.bd}

\cortext[cor1]{Corresponding author}
\affiliation[1]{
    organization={Department of Biomedical Engineering, Bangladesh University of Engineering and Technology},
    addressline={Dhaka-1205, Bangladesh}
}

\begin{abstract}
~\textbf{Objective: }Lung cancer is one of the leading causes of cancer-related deaths worldwide, and early detection is crucial to improve patient outcomes. However, early diagnosis is a major challenge, particularly in low-resource settings with limited access to medical resources and trained radiologists. This study aims to propose an automated end-to-end deep learning-based framework for the early detection and classification of lung nodules, specifically for low-resource settings. \textbf{Methods: }The proposed framework consists of three stages: lung segmentation using proposed 3D Res-U-Net, nodule detection using YOLO-v5, and classification with a Vision Transformer-based architecture. We evaluated the proposed framework on the publicly available dataset, LUNA16. The performance of the proposed framework was measured using the respective domain's evaluation metrics. \textbf{Results: }The proposed framework achieved a 98.82\% lung segmentation dice score while detecting the lung nodule with 0.76 mAP@50 from the segmented lung at a low false positive rate.  Furthermore, our proposed Vision transformer network achieved an accuracy of 96.29\%, which is 4.25\% higher than the state-of-the-art networks. The performance of all three networks in the proposed framework was compared with state of the art  studies, and they were found to be outperformed them in terms of evaluation metrics.\textbf{Conclusion: }Our proposed end-to-end deep learning-based framework can effectively segment the lung and detect and classify lung nodules, particularly in low-resource settings with limited access to radiologists. The proposed framework outperforms existing studies across all respective evaluation metrics. \textbf{Significance: }The proposed framework can potentially improve the accuracy and efficiency of lung cancer screening, ultimately leading to better patient outcomes, even in settings with limited medical resources and trained personnel.
\end{abstract}


\maketitle

\section{Introduction}
\par Lung cancer has the most significant mortality rate (18.7\%) and the second-highest incidence rate (12.4\%) among all malignant tumours worldwide, according to statistics from the GLOBOCAN 2022 cancer report~\cite{bray2024global}. Almost half (49\%) of all lung cancer cases now occur in countries ranked as medium to low on the Human Development Index (HDI)~\cite{cheng2016international}. Just 15\% of lung cancer patients survive five years after diagnosis, and 70\% have either locally progressed or distant metastases\cite{amin2017american}. The primary cause of this is that individuals with early-stage lung cancer frequently do not show any symptoms,  leading to a delayed diagnosis and worse outcomes. Pulmonary nodules are the main symptom of early-stage lung cancer, and early identification of lung cancer is essential for increasing survival rates\cite{chen2020decision,wu2018decision}. Accurately identifying lung nodules with a diameter between 3mm and 30mm is essential for early diagnosis and lowering mortality rates. Doctors use computed tomography (CT) imaging to detect pulmonary nodules and identify early-stage lung cancer. However, in developing nations, access to this technology is limited due to insufficient medical equipment and supplies. Additionally, CT exams are time-consuming and challenging, requiring radiologists to examine hundreds of images. This increases the likelihood of misdiagnosis. While rule based traditional automatic diagnosis of pulmonary nodules presents several challenges, such as the significant disparity between the number of positive and negative samples, the varying shapes of nodules, and the lack of integrated approaches \cite{wu2018decision,shrivastava2016training}.
\par
Recent surveys highlight that CNN architectures to transformer-based approaches can accelerate the computer-aided diagnosis (CAD); that can help address these issues by assisting doctors, speeding up diagnoses, and reducing the number of incorrect diagnoses.~\cite{khan2025cnn_survey,khan2023vit_survey,khan2023covid_survey,khan2023monkeypox_survey}. On the translational side, domain adaptation for CT infection assessment~\cite{owais2024meta} and attention–residual ViT designs for noise-robust classification~\cite{hameed2024arivit} reflect the shift toward transformer/hybrid models. A large body of infectious-disease imaging work established CNN baselines and boosted variants on chest X-ray/CT~\cite{khan2021covid_cnn,khan2021covid_boosted,khan2022covid_cbcnn,khan2022covid_cbcnn2,khan2023covid_boostedcnn}. Recently, transformer adaptations and SwinT-based feature-map enhancement reported for other conditions diagnosis(e.g., monkeypox, lung sound analysis)~\cite{khan2024rsfme,shuvo2025multi,shuvo2020lightweight} and biological process analysis~\cite{shuvo2025rescap}.
\par
Collectively, these works motivate a deployment-oriented, ROI-driven pipeline for lung CT; however, few studies connect lung parenchyma segmentation as a gating step, slice-level nodule detection, and ViT-based malignancy classification while also reporting error modes and deployment metrics.

\begin{figure*}[h]
    \centering
    \includegraphics[width=\linewidth]{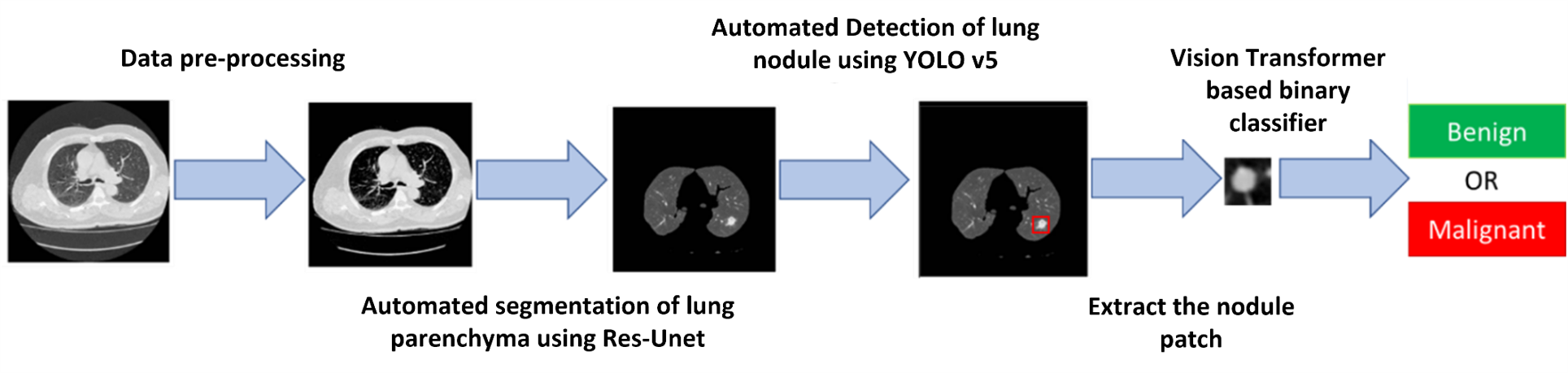}
    \captionsetup{justification=centering}
    \caption{A graphical overview of the end-to-end deep learning-based framework for lung cancer detection workflow. Original CT (lung window $-1200/600$ HU), 3D Res-UNet lung mask overlay, YOLOv5s detections (conf $\geq$0.40, NMS IoU=0.50), and the 64×64 classification patch centered on the detection for the ViT classifier.}
    \label{f1}
\end{figure*}

Building on the recent trends in transformer-based and hybrid approaches,this study presents a deep-learning model for automatically locating pulmonary nodules and determining whether they are malignant. The proposed model consists of three phases that identify and categorize pulmonary nodules. A 2D-based YOLOv5 network detects lung nodules on axial slices after the lung is segmented using a novel 3D Res-U-Net-based lung segmentation network. This process identifies patch representing the region of interest (ROI). The third and final phase is a 2D classification network that extracts nodule patch characteristics and forecasts their diagnostic score using a self-attention-based vision transformer. Unlike end-to-end pipelines that perform detection directly on full CT volumes, our framework introduces a dedicated lung segmentation stage. This step confines subsequent analysis to the lung parenchyma, effectively reducing irrelevant background, suppressing artifacts, and lowering false positives. Moreover, enabling clinicians to visually validate that only the lung region is analyzed. This modular design not only improves robustness across heterogeneous datasets but also facilitates transparent evaluation of each component in the pipeline.  A flow diagram of the proposed proposed model is shown in Figure~\ref{f1}.
\noindent The main contributions of this work are as follows:
\begin{enumerate}
    \item \textbf{End-to-end framework:} We design \textit{AutoLungDx}, a unified three-stage pipeline for lung CT analysis that integrates segmentation, detection, and classification in a clinically meaningful sequence (ROI gating $\rightarrow$ nodule detection $\rightarrow$ malignancy classification).
    
    \item \textbf{Segmentation innovation:} We propose a novel 3D Res-U-Net for lung parenchyma segmentation, which improves boundary accuracy (especially in juxta-pleural regions) over the baseline 3D U-Net by incorporating residual connections.
    
    \item \textbf{Detection stage:} We adapt YOLOv5s for nodule detection on axial CT slices, achieving competitive accuracy with efficient inference. We also analyze typical error patterns (small juxta-pleural misses and vessel-crossing false positives).
    
    \item \textbf{Classification stage:} We employ a Vision Transformer (ViT) for malignancy classification, demonstrating superior performance compared to conventional CNN-based baselines and capturing clinically relevant features in attribution maps.
    
    \item \textbf{Comprehensive evaluation:} We benchmark all stages on the LUNA16 dataset and show that the proposed framework achieves state-of-the-art performance across multiple tasks (Dice = 98.82\% for segmentation; mAP@50 = 0.76 for detection; ROC–AUC = 0.989 and Accuracy = 96.29\% for classification).
    
    \item \textbf{Deployment perspective:} Beyond accuracy, we assess inference latency on a moderate GPU system (3D Res-U-Net $\approx$1.6 s/sample; YOLOv5s $\approx$3.5–4.7 ms/sample; ViT $\approx$8–10 ms/sample), providing insights into clinical feasibility.
\end{enumerate}

\begin{table*}[h!]
\caption{Summary of Works Carried out in Segmentation, Nodule Detection and Classification Domain on Lung CT}
\label{tab:rworks}
\centering
\renewcommand{\arraystretch}{1.2}
\resizebox{\textwidth}{!}{
    \begin{normalsize}

\begin{tabularx}{\textwidth}{lccc}
\hline
\multicolumn{4}{c}{\textbf{Lung Segmentation Task}} \\
\hline
\hline
Author & Dataset & Network & Dice Score(\%) \\
\hline
Kendall et al. ~\cite{Badrinarayanan2017segnet} (2017) & LUNA16 & Segnet & $96.52$ \\
Alom et al.~\cite{alom2019recurrent} (2019) & LUNA16 & R2U-Net & $98.80$ \\
Khanna et al.~\cite{khanna2020deep} (2020) & LUNA16 & ResNet 34 based U-Net & $96.57$ \\
Jalali et al. ~\cite{jalali2021resbcdu} (2021) & LIDC-IDRI & BCDU-Net & $97.31$ \\
Francis et al. ~\cite{francis2022thoraxnet} (2022) & \parbox[c]{5cm}{\centering AAPM Lung CT Segmentation \\ Challenge 2017 dataset} & 3D U-Net & \parbox[c]{4cm}{\centering Left lung: $97.9$ \\ Right Lung: $97.3$} \\
Lijing et al. ~\cite{sun2024nodule} (2024) & LIDC-IDRI & U-Net & $88.17$ \\
&&& \\
\hline
\multicolumn{4}{c}{\textbf{Lung Nodule Detection Task}} \\
\hline
\hline
Author & Dataset & Network & mAP@50 \\
\hline
Cai et al. \cite{cai2020mask} (2020) & LUNA16 & Mask R-CNN & 0.88 \\
Shao et al. ~\cite{shao2021imageological} (2021) & LUNA16 & Yolo v5 refined & 0.75 \\
Zhou et al. \cite{zhou2022cascaded} (2022) & LUNA16 & Yolo v5(s) & 0.75 \\
Liu et al. \cite{liu2022stbi} (2022) & LUNA16 & STBi-YOLO & 0.95 \\
Liu et al. \cite{liu2025optimized} (2025) & LUNA16 & Mask R-CNN & 0.97 \\
&&& \\
\hline
\multicolumn{4}{c}{\textbf{Lung Nodule Malignancy Classification Task}} \\
\hline
\hline
Author & Dataset & Network & Accuracy(\%) \\
\hline
Filho et al. ~\cite{de2018classification} (2018) & LIDC-IDRI & CNN & $92.63$ \\
Xie et al. \cite{xie2018fusing} (2018) & LIDC-IDRI & DCNN & $89.53$ \\
Huang et al. \cite{huang2022self} (2022) & LIDC-IDRI & SSTL-DA & $91.07$ \\
Wu et al. \cite{wu2023self} (2023) & LIDC-IDRI & STLF-VA & $92.36$ \\
Lijing et al.
\cite{sun2024nodule} (2024) & LIDC-IDRI & Nodule-CLIP & $90.6$ \\
\hline
\hline
\end{tabularx}
    \end{normalsize}
    }
\end{table*}

The remainder of this paper is organized as follows. Section \ref{mt2} reviews related work in the field of lung cancer detection and highlights the strengths and limitations of existing approaches. Section \ref{mt3} presents utilized datasets and their preprocessing schemes. In Section \ref{deep}, we introduce the design of the proposed model in detail, including three architectures employed and the evaluation metrics used to assess the framework's performance.Section \ref{mt6} describes the experimental setup, evaluation metrics, and augmentation methods used in this study, \ref{mt7} presents the experimental results, and \ref{mt8} discusses the strengths and limitations of our proposed approach. Finally, Section \ref{mt10} summarizes the contributions of this work and outlines directions for future research.

\section{Related Works}\label{mt2}
As previously discussed, several studies have explored segmenting the lung parenchyma, identifying nodules, and categorizing malignant nodules using machine learning and deep learning-based frameworks, utilizing both publicly and privately available datasets. In this section, we provide an overview of research activities in this domain, as shown in Table  \ref{tab:rworks}).

\section {Materials and methods} \label{mt3}
In this section, we provide an overview of the dataset used and the signal preprocessing stages.
\subsection{Dataset}

The Lung Nodule Analysis (LUNA16) Dataset, which is a subset of the LIDC-IDRI dataset\cite{lidc2011x}, is used in this study. LIDC-IDRI contains 1018 low-dose CT scans and annotations from multiple radiologists. The case name, centroid coordinates of the nodule, diameter, volume of the nodule, texture of the nodule, and results of multiple radiologists are included in LIDC-IDRI. CT images of LIDc-IDRI with a slice thickness of more than 2.5 mm are eliminated from the LUNA16 dataset. This dataset was created to facilitate the development of automated nodule detection systems for clinical use as part of the 2016 International Symposium on Biomedical Imaging (ISBI) Lung Nodule Analysis (LUNA) Challenge. It consists of CT scans from 888 patients, with 1,186 nodules annotated by four experienced radiologists. To ensure accuracy, a two-phase annotation process was used, and nodules larger than 3mm were considered by all four radiologists~\cite{setio2017validation}. Therefore, nodules smaller than 3 mm or with only one or two radiologists' annotations are excluded. Thus, 888 low-dose CT scans are produced. It also includes a lung segmentation binary mask for each CT scan. This dataset is widely recognized as a benchmark for nodule detection research. It has been extensively used for training deep learning models and comparing the performance of different nodule detection algorithms. Malignancy markers were initially absent from the LUNA16 dataset. We used the metadata CSV file with the LIDC-IDRI dataset to update the radiologists' malignancy markings corresponding to patient IDs. Nodules are marked on a scale of 1 to 5 in the LIDC-IDRI dataset, with one denoting benign and 5 denoting the highest level of malignancy. According to the
malignancy score, the annotated nodule with a scoreless
then three is benign, a score equal to three is uncertain, and a
score greater than three is malignant.

\subsection{Data Prepossessing}
Lung CT scans typically display the entire chest cavity; however, pulmonary nodules are only visible in the left and right lobes. Occasionally, spherical tissues resembling nodules near the lobes may hinder accurate identification. To address this challenge, the raw CT data must be preprocessed, and the lung parenchyma needs to be extracted. This approach reduces the detection area and improves the accuracy of subsequent diagnoses.

The Hounsfield Unit (HU) is commonly used to represent CT data and is standardized. It provides a numerical value that indicates the radiodensity of tissues. It can vary across different tissues, leading to discrepancies when comparing regions of interest. This variation is primarily due to the composition and density of the tissues. By adjusting the HU value range, it is possible to control the detecting area and focus on the regions that are most relevant to the analysis, improving the accuracy of downstream tasks such as segmentation and detection.

\par The stages involved in data preparation are as follows:

\begin{itemize}
\item  \textbf{Data normalization :} The HU value range for lung parenchyma is between [-1200, 600]. Therefore, we normalize the CT images' HU values to fall within this range~\cite{tan2021data}. This ensures that all images are consistently represented, regardless of variations in acquisition protocols or scanners.

\item \textbf{Mapping:} The HU value range is mapped to [0, 1] using Min-Max normalization. This transformation makes the data suitable for machine learning models by scaling it to a standard range, facilitating better model convergence during training.

\item \textbf{Binarization of Mask:}
The provided ground truth mask contains several labels, with 1 representing the background and 3 or 4 representing the left and right lungs. We assign 0 to the background and 1 to both lung regions.

\item \textbf{Respacing and Resizing of Image and Mask:} 
We resize the data to $256\times256\times256$ and scale it to 1 mm in the x, y, and z dimensions.

\item  \textbf{Labeling Nodules:} We average the malignancy labels of the nodules and classify them as follows: if the mean malignancy level is less than three, the nodule is labelled as benign; if it is greater than three, the nodule is labelled as malignant; and if the mean malignancy level is equal to three, it is labelled as ambiguous.

\end{itemize}

To prepare data for YOLO-V5, the following steps are taken:

\begin{itemize}
\item  \textbf{Crop foreground:} We crop the space around the segmented lung parenchyma based on the largest foreground region.
\item  \textbf{Generate bounding box:} We use the spherical mask to generate a bounding box for each slice based on the provided diameter and center from the annotation file. The YOLOv5 label format includes the class ID, x-center, y-center, width, and height, with normalized coordinates.

\end{itemize}
\par
To train the classification network, a $64\times64$ slice is cropped from the spherical mask volume according to the diameter and center of the annotation file.

\section{Deep learning architectures}\label{deep}

\subsection{Segmentation Network}
Lung CT scans typically display the entire chest cavity of the patient, offering a comprehensive view of the thoracic region. These scans are vital for diagnosing a range of pulmonary conditions. However, they often present challenges in interpretation due to the presence of various structures within the chest, including several spherical tissues adjacent to the lobes. These structures can appear similar to nodules, complicating the analysis and detection process. These misclassifications can lead to false positives or missed diagnoses, making accurate detection of actual nodules difficult. Therefore, it is essential to preprocess the raw CT data to enhance the accuracy of the analysis. This preprocessing involves delineating the lung parenchyma and removing surrounding tissues that may obscure the actual nodules. By isolating the lung parenchyma from other structures, we reduce the area the model needs to process, allowing for more accurate segmentation and detection of nodules.

\par
\begin{figure}[h]
    \centering
    \includegraphics[width=\linewidth]{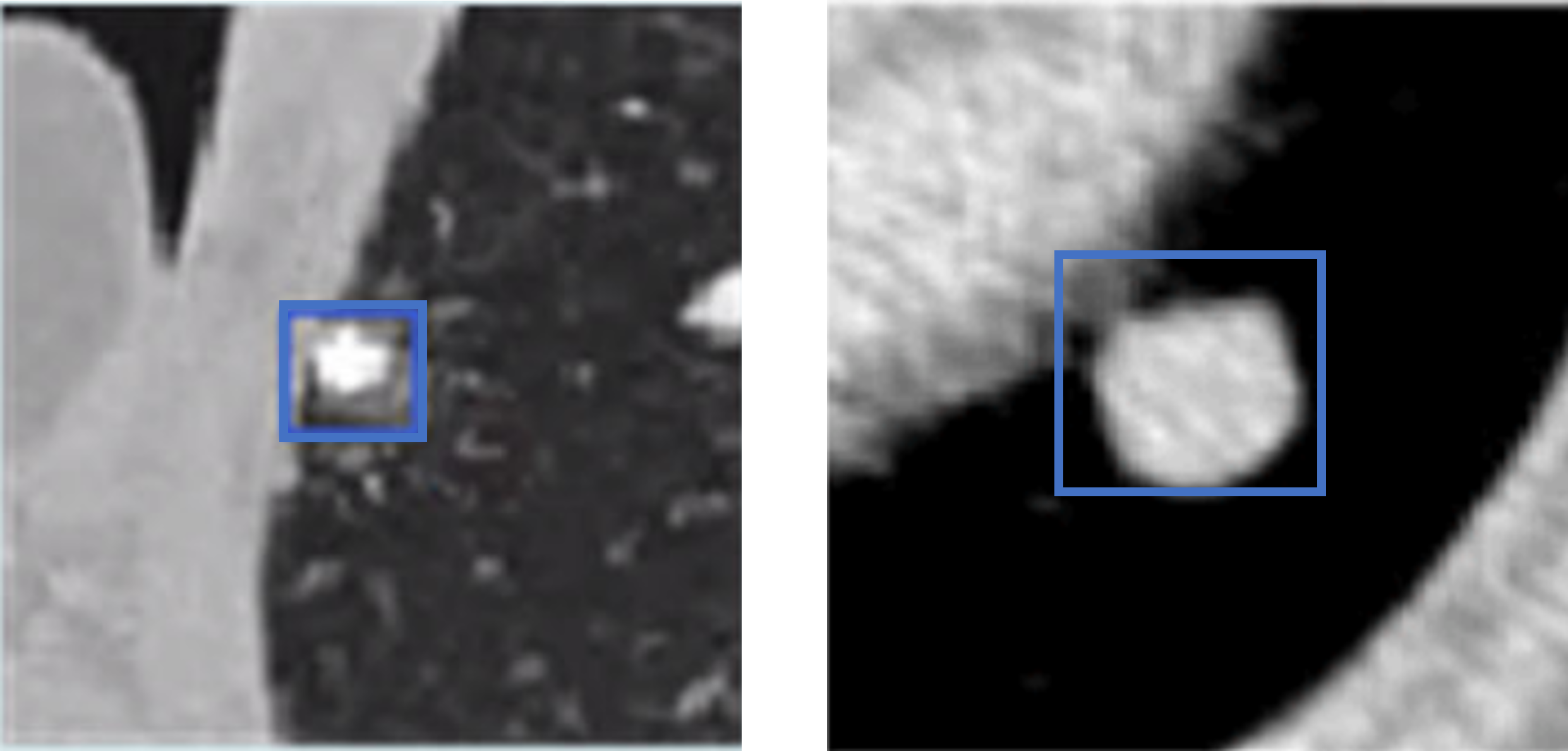}
    \caption{Some instances of nodules very near the lung surface.}
    \label{Pic9}
\end{figure}

  Additionally, different types of nodules can be present in the lung, some of which are located very close to the lung surface, as shown in Figure~\ref{Pic9}. Therefore, maintaining a reliable and precise segmentation network is crucial. This is why we have chosen to employ 3D methods in our approach, as they have demonstrated significantly better performance in lung segmentation tasks than traditional 2D methods. By utilizing 3D models, we are able to capture spatial relationships within the lung tissue more effectively, allowing the model to consider the whole 3D structure of the lungs and improve its accuracy in segmenting the nodule areas.
\begin{figure*}[h!]
    \centering
    \includegraphics[width=\linewidth]{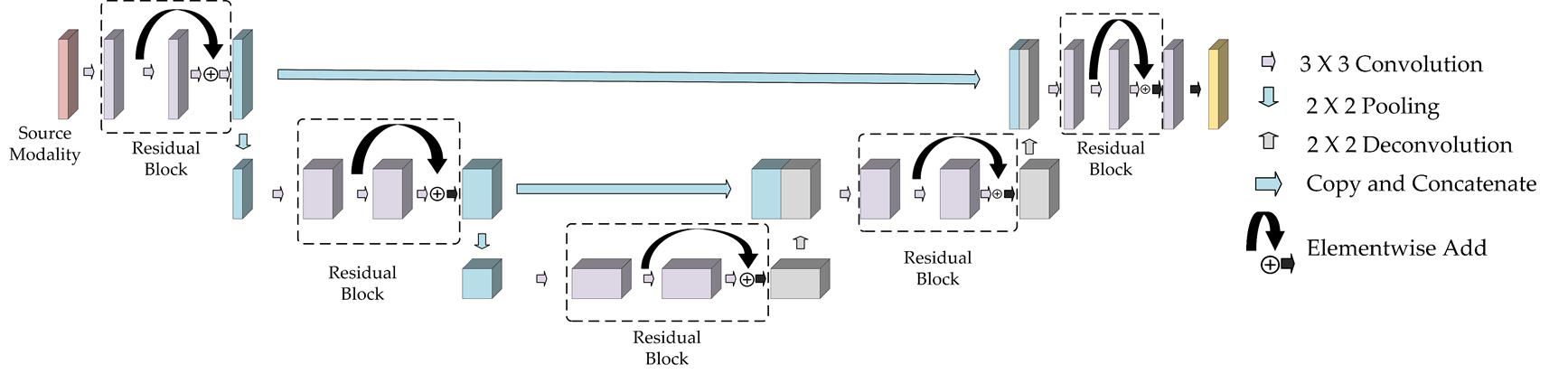}
    \caption{Overview of the proposed 3D Res-Unet architecture}
    \label{modelS}
\end{figure*}

\subsubsection{Baseline Model-3D Unet }
Comparable to the standard U-Net \cite{ronneberger2015u}, our 3D U-Net model consists of an analysis and synthesis path with four resolution stages, with the number of filters being 24, 48, 96, and 192, respectively. Each layer in the analysis path contains two $3\times3\times3$ convolutions, followed by a rectified linear unit (ReLU) activation function and a $2\times2\times2$ max pooling operation with two strides in each dimension. In the synthesis path, each layer comprises a $2\times2\times2$ up-convolution with two strides in each dimension, followed by two $3\times3\times3$ convolutions, each followed by a ReLU activation function.

Shortcut connections from layers of the exact resolution in the analysis path are used to provide the synthesis path with the necessary high-resolution features. In the final layer, a $1\times1\times1$ convolution reduces the output channels to the number of labels, in this case, 3. The input to the network is a $256\times256\times256$ voxel image tile, and the output from the last layer is $256\times256\times1$ voxels in the x, y, and z directions, respectively.

\subsubsection{Proposed Model (3D Res-Unet) }
The most crucial component of a deep residual network is the residual block, which consists of convolutional layers with identity mapping, as shown in Figure~\ref{modelS}. The residual block is designed to mitigate the vanishing gradient problem and may also enhance information transmission between layers. In this study, four subunits (four convolutional layers within the residual block) were used.
\subsection{Detection Network}
\begin{figure}[t]
    \centering
    \includegraphics[width=0.98\linewidth]{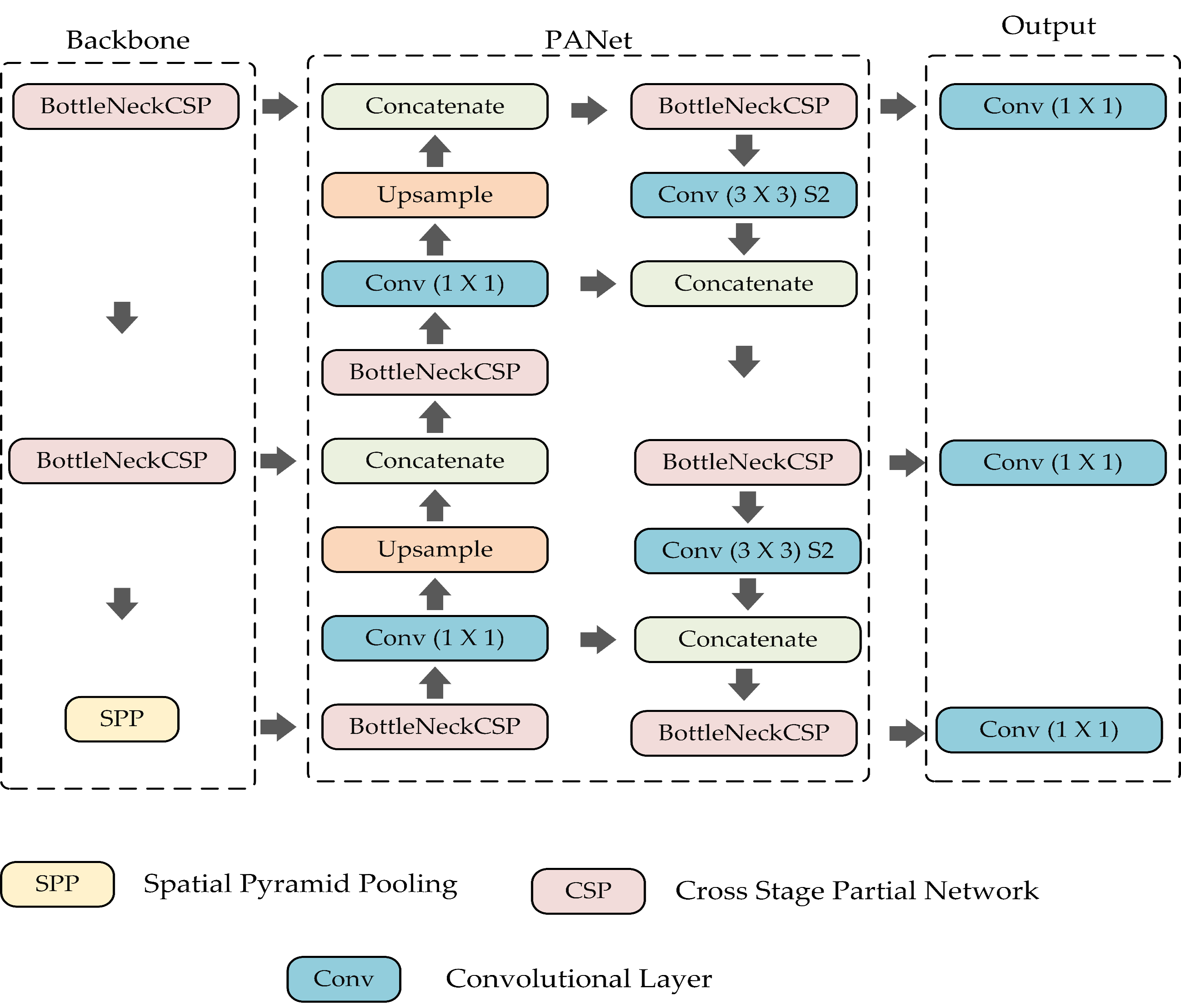}
    \caption{The main components of the YOLOv5 model include the CSPDarknet backbone, the PANet neck, and the YOLO Layer head. In the CSPDarknet, features are extracted from the input data. These extracted features are then combined in the PANet. Finally, the YOLO Layer generates the object detection results, which include the class, score, location, and size of the detected objects.}
    \label{model2}
\end{figure}

YOLOv5 is a state-of-the-art deep learning model for object detection that utilizes a single neural network to predict object bounding boxes and class probabilities in an input image. This model improves upon its predecessors, YOLOv4 and YOLOv3, by incorporating several new techniques to achieve higher accuracy and efficiency. YOLOv5 features a CSPNet (Cross-Stage Partial Network) backbone, an efficient variant of the Darknet backbone used in previous YOLO versions. The CSPNet backbone extracts features from the input image, which are then processed through multiple layers of spatial pyramid pooling (SPP) and path aggregation network (PAN) modules, enhancing the model's ability to capture features at different scales, as shown in Figure~\ref{model2}.

The YOLOv5 model utilizes anchor boxes to predict object bounding boxes and incorporates a novel loss function called the Generalized Intersection over Union (GIoU) loss for training. The GIoU loss enhances the traditional Intersection over Union (IoU) loss, providing a better measure of the similarity between predicted and ground truth bounding boxes. Additionally, YOLOv5 employs a self-ensemble technique, where multiple versions of the same model make predictions on the same input image. The predictions from these models are then combined to generate a final prediction, which has been shown to improve accuracy~\cite{yolov5}. In this study, we use the small version of this model named YOLOv5s due to computational constraints.

\subsection{Classification Network}

The Vision Transformer (ViT) is a deep learning architecture for computer vision tasks, introduced by Google researchers in 2020~\cite{dosovitskiy2020image}. Unlike traditional CNNs, ViT utilizes self-attention mechanisms from the transformer architecture to capture global context from images, enabling the model to focus on important features across the entire image rather than local regions. This self-attention mechanism allows ViT to model long-range dependencies, which is especially useful for large, complex images.

\begin{figure}[h]
    \centering
    \includegraphics[width=0.98\linewidth]{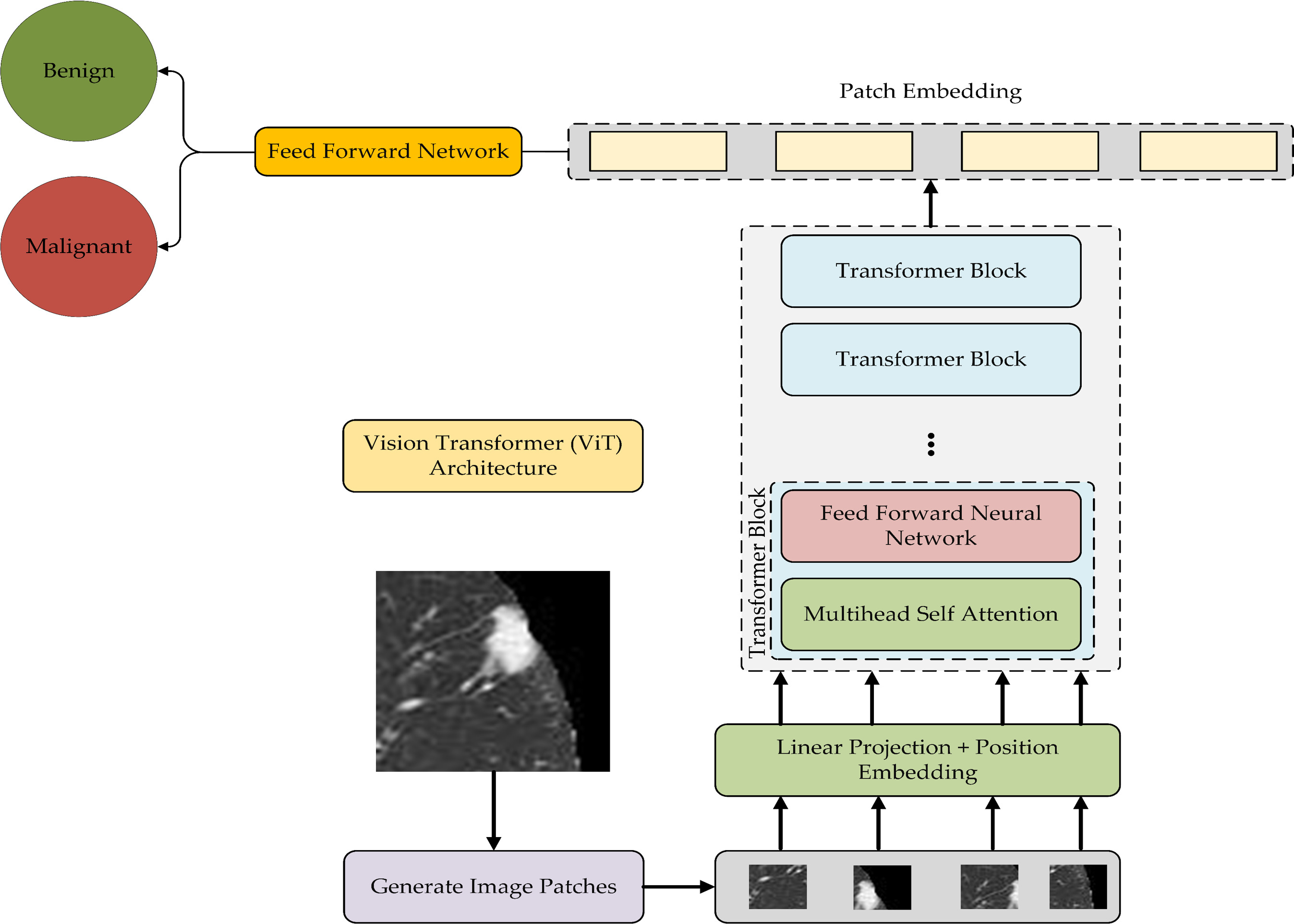}
    \caption{Overview of the proposed vision transformer architecture.}
    \label{vit}
\end{figure}

In our proposed ViT network, the input nodule patches are first divided into eight smaller patches. These patches are then subjected to a series of linear transformations, converting them into a sequence of tokens. The tokens are passed through 8 successive transformer encoders, where multi-head self-attention with a projection dimension of 64 and feed-forward layers with 64 and 128 hidden nodes are applied to extract image features. These encoder layers are designed to learn complex relationships between patches, enhancing the model's ability to capture detailed contextual information. Finally, a feed-forward network with sequential hidden states of 2048 and 1024 helps refine the learned features, assisting the final classification layer in predicting the output label. This approach facilitates the model's ability to generalize well, offering accurate predictions across various imaging conditions.

\section{Experimental Evaluation} \label{mt6}
\subsection{Experimental Setup}
The deep learning architectures are implemented using PyTorch, TensorFlow, and Keras, with all models trained and tested on an Intel® Xeon® CPU and an NVIDIA GTX 2080 Ti GPU. The models were trained on 4 RTX A4500 GPUs, providing a total of 80 GB of GPU memory, with 24 GB of VRAM per GPU. The system also includes an Intel Core i9-10920X CPU with 12 cores, a clock speed of 3.50 GHz, and 256 GB of RAM.

\par 
Details regarding the size of training and testing samples for each task, along with all hyperparameters used for training, are provided in Table~\ref{tab:hyper}. During training, 10\% of the training data was reserved for validation. Furthermore, a mini-batch balancing strategy was employed for classification model training to ensure each batch contained an equal number of samples from each class.

\subsection{Evaluation Metrics for Lung Segmentation}
The Dice score is a widely used metric for evaluating the performance of lung segmentation algorithms, as it measures the pixel-wise agreement between the predicted lung segmentation and the corresponding ground truth. The Dice score is calculated as follows:

\begin{equation}
    \text{Dice Score} = \frac{2 \times |X \cap Y|}{|X| + |Y|}
\end{equation}

In this formula, $X$ and $Y$ represent two sets, and $|X|$ and $|Y|$ represent their respective cardinalities (i.e., the number of elements in each set). The symbol $\cap$ represents the intersection of the two sets (i.e., the set of elements that are common to both $X$ and $Y$), and $|X \cap Y|$ represents the cardinality of their intersection. The dice score is a measure of similarity between two sets, with a value ranging from 0 (no similarity) to 1 (perfect similarity).
\par
Another commonly used evaluation metric is the Intersection over Union (IoU), also known as the Jaccard Index, which quantifies the overlap between the predicted segmentation and the ground truth relative to their union. IoU is defined as:

\begin{equation}
    \text{IoU} = \frac{|X \cap Y|}{|X \cup Y|}
\end{equation}

Here, $X \cup Y$ denotes the union of the two sets, and $|X \cup Y|$ represents its cardinality. Like the Dice score, IoU ranges from 0 to 1, where 0 indicates no overlap and 1 indicates perfect overlap. While Dice emphasizes the intersection relative to the average size of the sets, IoU emphasizes the intersection relative to the total combined region, making both metrics complementary in assessing segmentation performance.

\subsection{Evaluation Metrics for Lung Nodule Detection}
To evaluate the performance of the YOLOv5 nodule detection network, two metrics, mAP@50 and mAP@50:95, have been utilized. mAP@50 refers to the mean average precision at a detection threshold of 50\% for the official PASCAL VOC (Visual Object Classes) dataset. It is calculated as follows:

\begin{equation}
mAP@50 = \frac{1}{N_{classes}} \sum_{i=1}^{N_{classes}} AP_{50}^{i}
\end{equation}

where $N_{classes}$ is the number of object classes, and $AP_{50}^{i}$ is the average precision for class $i$ at a detection threshold of 50\%.

mAP@50:95 is the mean average precision averaged over all recall values between 50\% and 95\% for the official COCO (Common Objects in Context) dataset. It is calculated as follows:

\begin{equation}
mAP@50:95 = \frac{1}{N_{classes}} \sum_{i=1}^{N_{classes}} \int_{0.5}^{0.95} AP(r) \mathrm{d}r
\end{equation}

where $N_{classes}$ is the number of object classes, $AP(r)$ is the average precision at recall level $r$, and the integral is taken over all recall levels from 50\% to 95\%.

\begin{table*}[t]
\centering
\renewcommand{\arraystretch}{1.2}
\caption{Details of Hyperparameters Used for the Models}
\label{tab:hyper}
\resizebox{\textwidth}{!}{
\begin{normalsize}

\begin{tabular}{p{3cm}p{2cm}|p{3cm}p{4.5cm}|p{2.5cm}p{3.5cm}}
\hline
\multicolumn{2}{c|}{\textbf{Lung Segmentation Task}} &
\multicolumn{2}{c|}{\textbf{Lung Nodule Detection Task}} &
\multicolumn{2}{c}{\textbf{Lung Nodule Malignancy Classification Task}} \\
\hline\hline
\textbf{Hyper-parameters} & \textbf{Values} & 
\textbf{Hyper-parameters} & \textbf{Values} & 
\textbf{Hyper-parameters} & \textbf{Values} \\
\hline\hline
Training data & $800$ & Training data & $6045$ & Training data & $1577$ \\
Test data & $88$ & Test data & $756$ & Test data & $394$ \\
Batch size & $4$ & Batch size & $64$ & Batch size & $256$ \\
Learning rate & $0.0001$ & Learning rate & $0.01$ & Learning rate & $0.0001$ \\
Epoch & $60$ & Epoch & $600$ & Epoch & $60$ \\
Optimizer & Adam & Optimizer & Adam & Optimizer & Adam \\
\hline
 & & 
 & Bounding box regression loss & 
&  \\
Loss function& Dice loss  & Loss function & Objectness loss & Loss function  & Cross-entropy \\
 &  &  & Classification loss &  &  \\
\hline\hline
\end{tabular}
\end{normalsize}}
\end{table*}
\subsection{Evaluation Metrics for Cancer Classification}
The cancer classification LC-ViT framework is evaluated quantitatively using well-known and important performance evaluation metrics such as Accuracy, Precision, Recall, and F1-score. The formulas for these metrics are as follows:

Accuracy(Acc.) measures the proportion of correctly classified samples and is calculated as follows:
\begin{equation}
\\Acc.=\frac{TP+TN}{TP+FP+TN+FN}\
\end{equation}

Precision measures the proportion of positive samples that are correctly identified and is calculated as follows:
\begin{equation}
\\Precision=\frac{TP}{TP+FP}\
\end{equation}
Recall measures the proportion of positive samples that are correctly identified out of all the positive samples and is calculated as:
\begin{equation}
\\Recall=\frac{TP}{TP+FN}\
\end{equation}
F1-score combines both precision and recall into a single metric and is calculated as the harmonic mean of precision and recall:
\begin{equation}
\\F1\;score=\frac{2*TP}{2*TP+FP+FN}\
\end{equation}
Where TP, TN, FP, and FN represent the number of true positives, true negatives, false positives, and false negatives, respectively.

We also assessed threshold-free discrimination using Receiver Operating Characteristic (ROC) and Precision- Recall (PR) curves computed from the model’s probabilistic outputs on the held-out test set. ROC characterizes the trade-off between true-positive and false-positive rates across all thresholds; PR emphasizes performance under class imbalance by summarizing precision as a function of recall.

\subsection{Data Augmentation}
To improve model generalization and account for clinical variability, we applied task-specific data augmentation strategies. For lung segmentation, we used random flips along left–right and anterior–posterior axes to leverage lung symmetry, random rotations $\pm(10^\circ\text{–}15^\circ)$
 to handle orientation variability, affine and elastic deformations to simulate patient motion and scanner differences, and random cropping/zoom to accommodate variable fields-of-view. For lung nodule detection using YOLOv5s, we employed the model’s built-in augmentations with hyperparameters translate=0.1, scale=0.5, fliplr=0.5, mosaic=1.0, and other standard settings. For lung nodule malignancy classification using a ViT, we applied RandomFlip ("horizontal"), RandomRotation, and RandomContrast. These augmentations were designed to simulate realistic variations in patient positioning, image acquisition, and lung appearance, enhancing model robustness.

\section{Experimental results}  \label{mt7}

\subsection{Lung Segmentation Results}

We compared the results of our lung segmentation with the ground-truth lung masks provided by experts to evaluate the effectiveness of the proposed segmentation method. This comparison was necessary to assess how well the model performs in real-world applications, where accuracy and precision are critical. To quantify the comparison, we employed two distinct metrics: Dice score (DSC) and F1-score. To further examine the impact of residual connections on performance, we evaluated the performance of two models, 3D-U-Net and 3D-Res-U-Net, a modified version of 3D-U-Net that incorporates residual connections to improve information flow and gradient propagation. 

 \begin{figure*}[h!]
    \centering
    \includegraphics[width=0.8\linewidth]{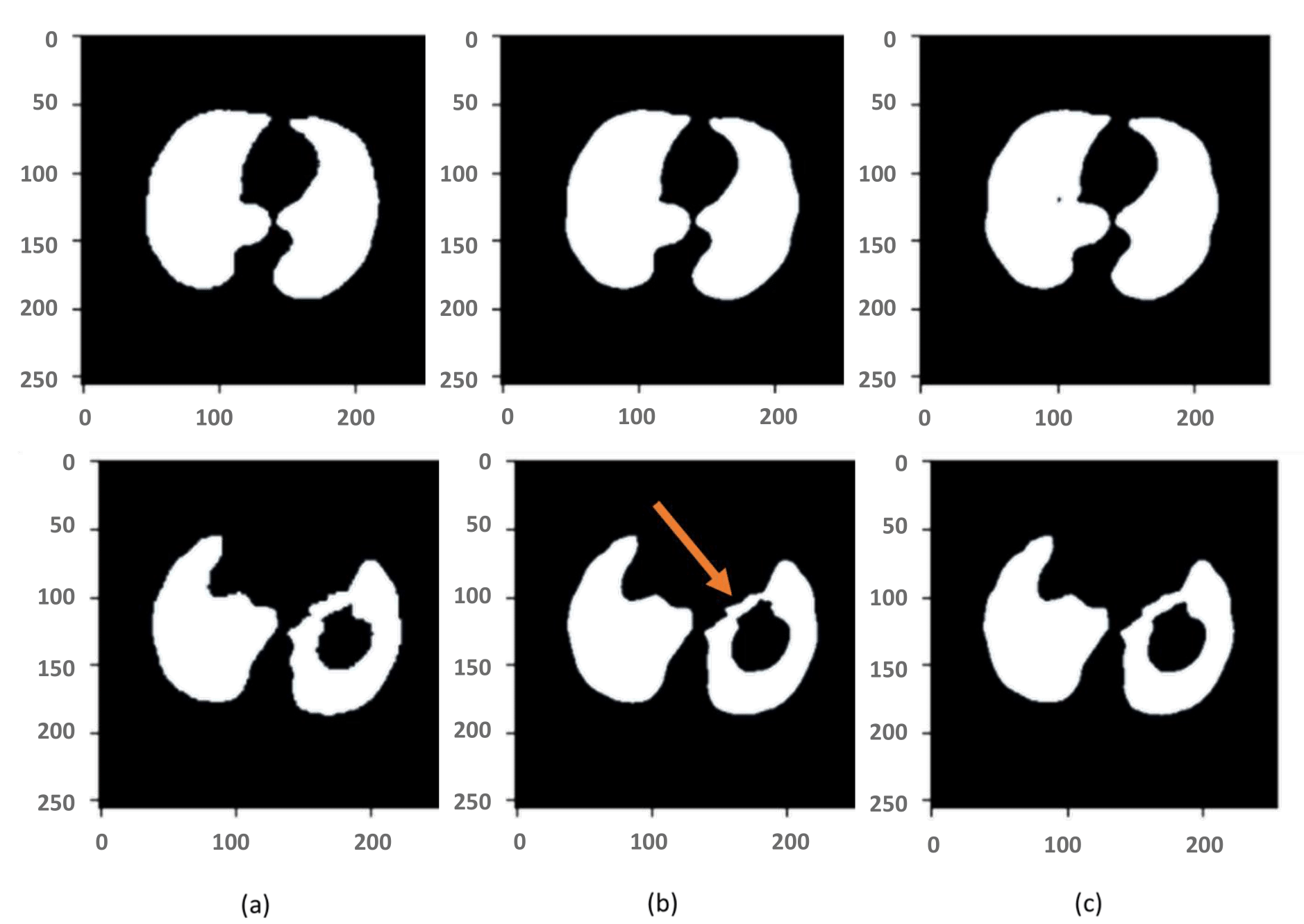}
\caption{Visual comparison of lung segmentation results obtained using the 3D U-Net and 3D Res-U-Net architectures. Column (a) shows the ground truth masks, column (b) displays the 3D U-Net predictions with Dice scores of 0.9464 (first row) and 0.9409 (second row), and column (c) presents the 3D Res-U-Net predictions with Dice scores of 0.9477 (first row) and 0.9505 (second row). The orange arrow highlights a typical juxta-pleural leakage observed in (b), which is effectively corrected in (c).}

    \label{s}
\end{figure*}

Table~\ref{segment} presents the overall performance of these different architectures. The proposed 3D Res-U-Net achieved a Dice score of 98.82\%, with an estimated 95\% CI of 98.5--99.1\%, indicating stable performance across folds, indicating that our model performed at a high level of accuracy. The 3D-Res-U-Net model highlights the proposed method's ability to generalize well to unseen samples, showing its robustness in segmenting lung structures across a range of input images. 

\begin{figure}[h!]
    \centering
     \includegraphics[width=1\linewidth]{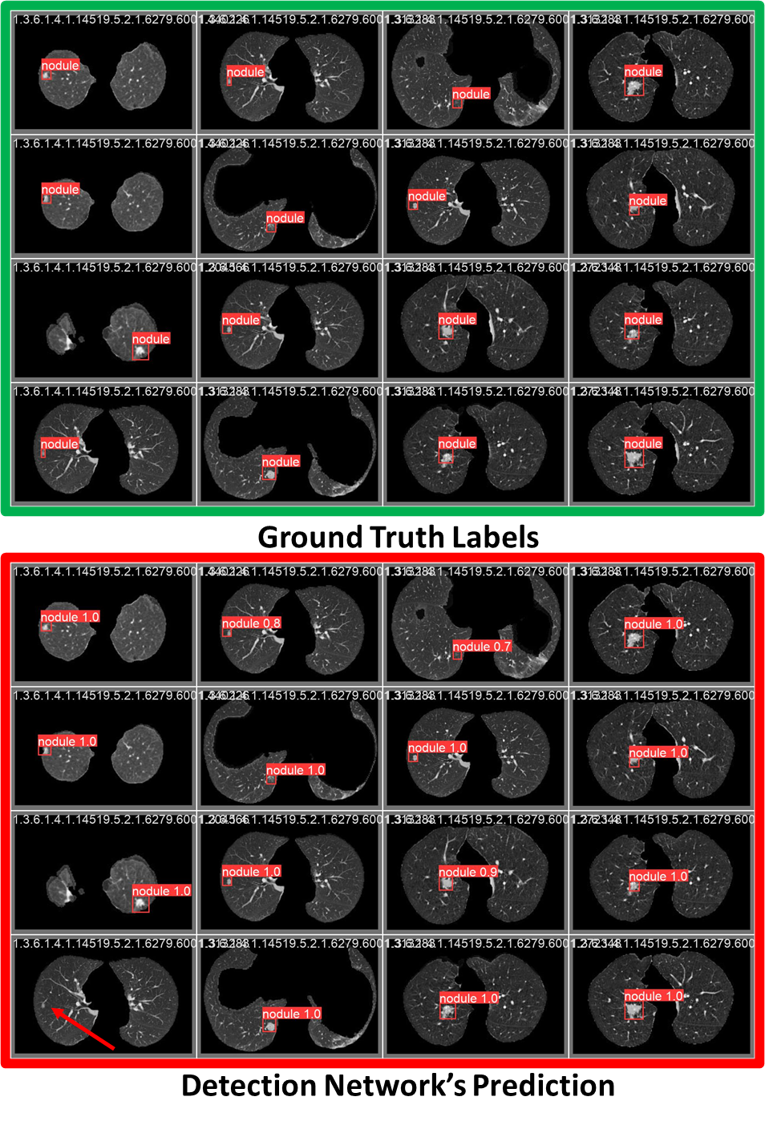}
    \caption{Detection results of nodules on the LUNA16 dataset using YOLOv5(s).Top (green frame): ground-truth annotations. Bottom (red frame): model predictions with confidence scores. Misses are enriched in small (3--6\,mm) juxta-pleural nodules; false positives often occur near vessel crossings or fissures.}
    \label{d}
\end{figure}


\begin{table*}[h!]
    \caption{Comparison of the Proposed Framework with Existing Lung Segmentation Methods}
    \label{segment}
    \centering
    \renewcommand{\arraystretch}{1}
     \scalebox{1}{
 \resizebox{\linewidth}{!}{
\begin{normalsize}

    \begin{tabular}{lccccc}
        \toprule
        \toprule
        \textbf{Author} & \textbf{Network} & \textbf{Dice (\%)} & \textbf{F1-Score (\%)} & \textbf{IoU (\%)} \\
        \midrule
        \midrule
        Kendall et al.~\cite{Badrinarayanan2017segnet} & SegNet & 96.52 & 95.75 & 94.0 \\
        Alom et al.~\cite{alom2019recurrent} & R2U-Net & 98.80 & \textcolor{red}{98.79} & 97.6 \\
        Lijing et al.~\cite{sun2024nodule} & U-Net & 88.17 & -- & 79.0 \\
        Jalali et al.~\cite{jalali2021resbcdu} & Res-BCDU-Net & 97.31 & 98.05 & 95.0 \\
        Rikhari et al.~\cite{rikhari2024fully} & Unet-Inception-ResNet-v2 & 97.13 & 97.01 & 94.8 \\
        \multirow{2}{*}{\textbf{Our proposed}} & 3D U-Net & 97.97 & 97.94 & 96.1 \\
        & 3D Res-U-Net & \textcolor{red}{98.82} & 98.73 & 97.6 \\
        \bottomrule
        \bottomrule
    \end{tabular}
    \end{normalsize}}}
\end{table*}

Additionally, the 3D-Res-U-Net outperformed all previous state-of-the-art lung segmentation networks. To ensure that neither overfitting nor underfitting occurred, we also analyzed the learning curve of the proposed approach for lung CT segmentation in terms of dice loss.

We also evaluated the visual outcomes of the proposed method, and Figure~\ref{s} shows that there was minimal to no error in generating the lung segmentation mask from the input lung CT images. The segmentation process accurately delineated the lung boundaries, capturing even the subtle details in the lung parenchyma. The 3D-Res-U-Net effectively handled the most challenging regions of the lung due to its superior performance, particularly in areas with complex structures. In contrast, the 3D-U-Net struggled to reproduce the complex curvatures of the lung boundaries, as indicated by the red arrow in Figure~\ref{s}. This further emphasizes the advantage of the residual connections in the 3D-Res-U-Net, which facilitate more accurate feature learning and improve segmentation results in difficult-to-detect areas.

\subsection{Nodule Detection Results}

We evaluated YOLOv5(s) alongside other common nodule detection strategies from recent publications to assess its efficacy in single-stage lung nodule detection. This comparison is crucial for determining how well our proposed method performs relative to other well-established models. Table~\ref{Detect} presents the results of this comparison between our model, YOLOv5, YOLOv5s, and Mask R-CNN. For detection, the YOLOv5(s) model achieved an mAP@50 of 0.76 (mAP@50:95 = 0.62), with an estimated 95\% CI of 0.73--0.79. When comparing YOLOv5s to the single-stage SSD and the two-stage Mask R-CNN, it can be concluded that YOLOv5s is a more compact and efficient network model. Despite the reduced size, YOLOv5s achieves competitive performance while requiring fewer computational resources. The YOLOv5 model we used consistently performed well compared to other state-of-the-art architectures, demonstrating its robustness in nodule detection. However, STBi-YOLO demonstrated significantly better performance across all models, highlighting its potential for more accurate detection. We also observed that the preprocessing techniques and training data preparation process were not well detailed in \cite{liu2022stbi}, which may be a factor influencing the performance discrepancy between our model and others.

\begin{table}[h]
 \caption{Comparison of the Proposed Framework with Existing Nodule Detection Methods}
 \label{Detect}
 \centering
 \renewcommand{\arraystretch}{1.2}
 \scalebox{1}{
 \resizebox{\linewidth}{!}{
    \begin{normalsize}

 \begin{tabular}{llll}
\hline
 \hline
 
Author& Network & mAP@50 & mAP@50:95  \\
 \hline
 \hline
  Cai et al. \cite{cai2020mask} & Mask R-CNN &   $0.88$  & $0.571$  \\
  Shao et al. ~\cite{shao2021imageological} & Yolo v5 refined          &$0.75$ & -\\
  Zhou et al.\cite{zhou2022cascaded} & Yolo v5(s) & $0.75$ & -       \\
  Liu et al. \cite{liu2022stbi}   & STBi-YOLO    & \textcolor{red}{$0.95$} &- \\
  Our Proposed & YOLOv5(s) & $0.76$ &  \textcolor{red}{$0.62$}\\
\hline
\hline
 \end{tabular}
 \end{normalsize}
 }}
 \end{table}
 
We also examined the visual outcomes of the proposed method, and Figure\ref{d} shows that there was minimal to no error in detecting the lung nodule from the segmented lung CT images. YOLOv5 effectively handled the most challenging nodules, though smaller or less-represented classes were occasionally missed by the detection algorithm, as indicated by the red arrow in Figure\ref{d}. Therefore, YOLOv5 demonstrated its ability to work well on complex cases, including those with irregular shapes and sizes.

 \
\subsection{Cancer Classification Results}
\begin{figure}[h!]
    \centering
    \includegraphics[width=1\linewidth]{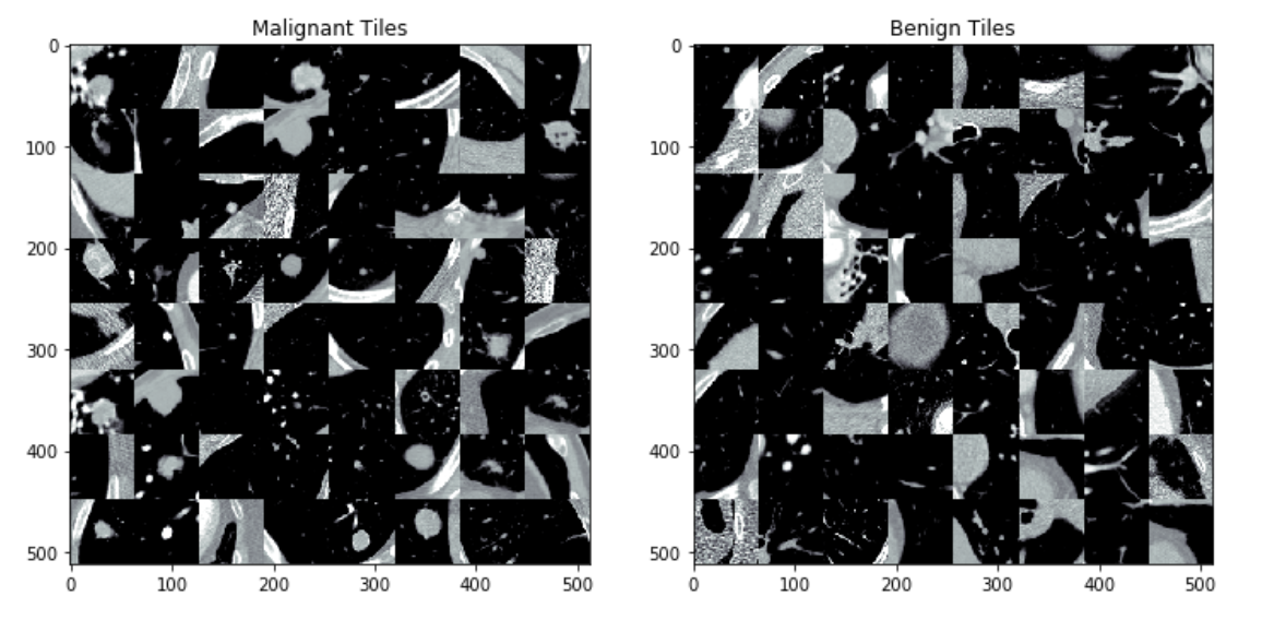}
    \caption{Representation of some malignant and benign nodules}
    \label{c}
\end{figure}

We evaluated our proposed Vision Transformer (ViT) algorithm on the LUNA16 dataset and collected diagnostic information from the LIDC-IDRI lung nodule dataset~\cite{armato2011lung}, which serves as a primary source of the LUNA16 dataset. This combination of datasets allows us to assess the model's generalizability and robustness across different lung CT images. The LIDC-IDRI dataset provides richly annotated nodules (examples of binary-class nodules from the dataset are shown in Figure\ref{c}), including malignancy ratings, which enhances the reliability of the evaluation process. 

Table\ref{tab:classification_comparison} compares the performance metrics of our proposed model with those of several state-of-the-art (SOTA) methods. It is important to note that all the studies referenced utilized the LIDC-IDRI database, making it a widely recognized benchmark in lung nodule detection. As shown in the table, our ViT model demonstrates promising performance, outperforming many existing methods. For instance, Wu et al. used STLF-VA classifiers and reported accuracies of 92.36\%. In comparison, our model achieved an accuracy of 96.29\%(95\% CI: 95.23–97.35\%) and a precision of 90.95\%, which represents an improvement of 4.25\% in accuracy over the previous top-performing method~\cite{wu2023self}. This improvement highlights the effectiveness of our ViT architecture in achieving high performance on a complex and challenging task like lung nodule detection.

\begin{table*}[h!]
\centering
\caption{Comparison of the proposed framework with existing lung nodule malignancy classification methods.}
\label{tab:classification_comparison}
\renewcommand{\arraystretch}{1}
\resizebox{\linewidth}{!}{
\begin{normalsize}
\begin{tabular}{lccccc}
\hline
\hline
\textbf{Author} & \textbf{Network} & \textbf{Acc. (\%)} & \textbf{Rec. (\%)} & \textbf{Prec. (\%)} & \textbf{F1-score (\%)} \\
\hline
\hline
Filho et al.~\cite{de2018classification} & CNN        & 92.03 & 93.47 & 90.70 & 92.07 \\
Xie et al.~\cite{xie2018fusing}           & DCNN       & 89.93 & 92.02 & 84.19 & 87.95 \\
Huang et al.~\cite{huang2022self}        & SSTL-DA    & 91.07 & 91.22 & 90.93 & 91.08 \\
Wu et al.~\cite{wu2023self}              & STLF-VA    & 92.36 & 93.08 & 90.93 & 91.99 \\
Lijing et al.~\cite{sun2024nodule}&Contrastive learning& 90.60& 92.60 &-&-\\
Gupta et al.~\cite{gupta2024texture}&ML& 91.30& 92.10 &-&-\\ 
Madhuri et al.~\cite{madhuri2025lmlcc}& Semi-Supervised DL &91.96 &92.24 &-&-\\
\textbf{Our Proposed}                    & \textbf{ViT} & \textbf{96.29} & \textbf{95.84} & \textbf{96.26} & \textbf{96.04} \\
\hline
\hline
\end{tabular}
\end{normalsize}
}
\end{table*}

\begin{figure}[!t]
  \centering
  \includegraphics[width=0.62\linewidth]{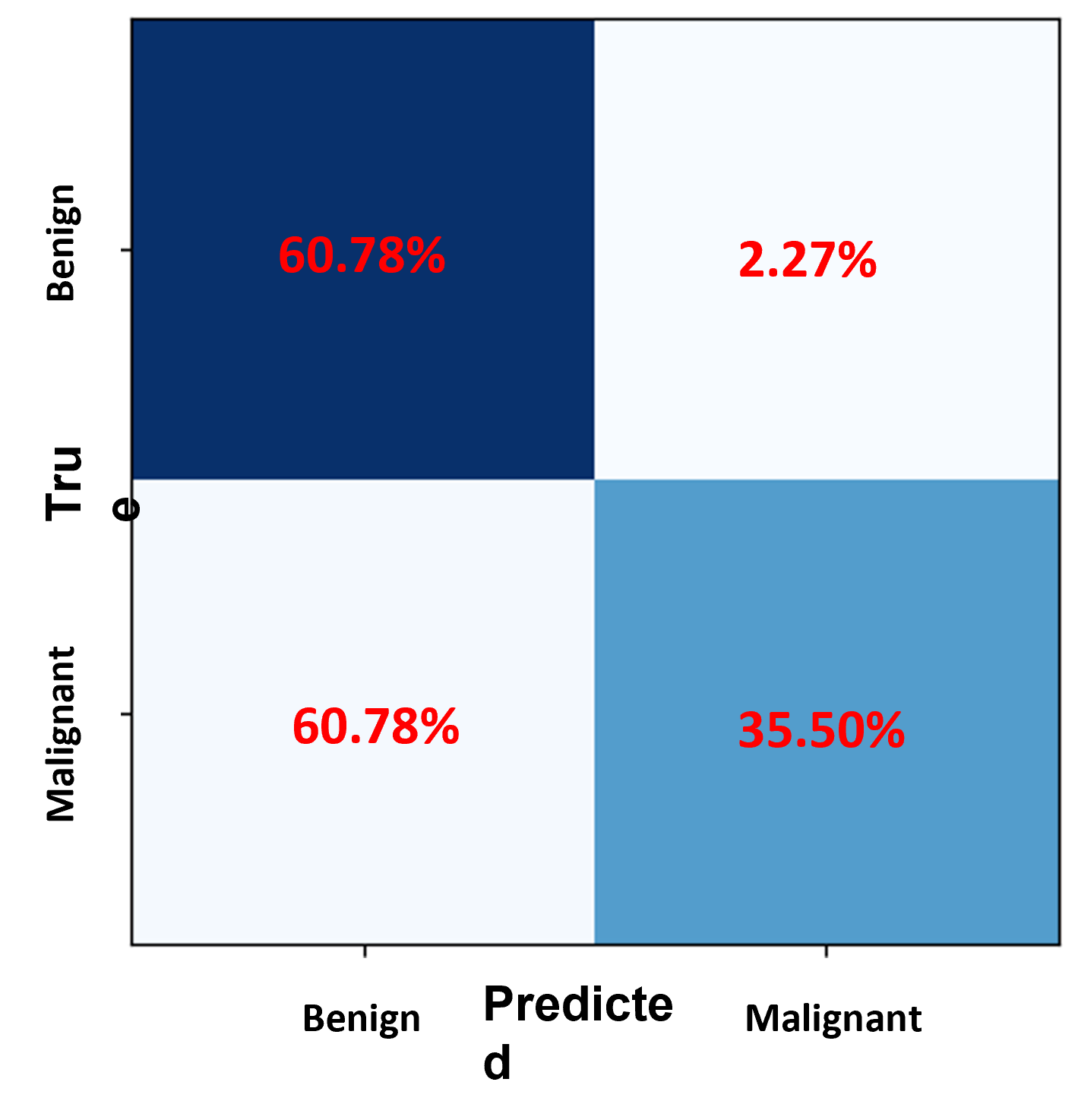}
  \caption{Confusion matrix on the held-out test set .}
  \label{fig:cm}
\end{figure}

The confusion matrix (Figure~\ref{fig:cm}) and Table~\ref{tab:classwise} shows our model yielding Accuracy 0.963, Balanced Accuracy 0.958, Sensitivity 0.940, Specificity 0.977, PPV 0.961, and NPV 0.964. Errors are mainly false negatives in small/low-contrast juxta-pleural nodules.

\begin{figure}[h!]
    \centering
    \includegraphics[width=1\linewidth]{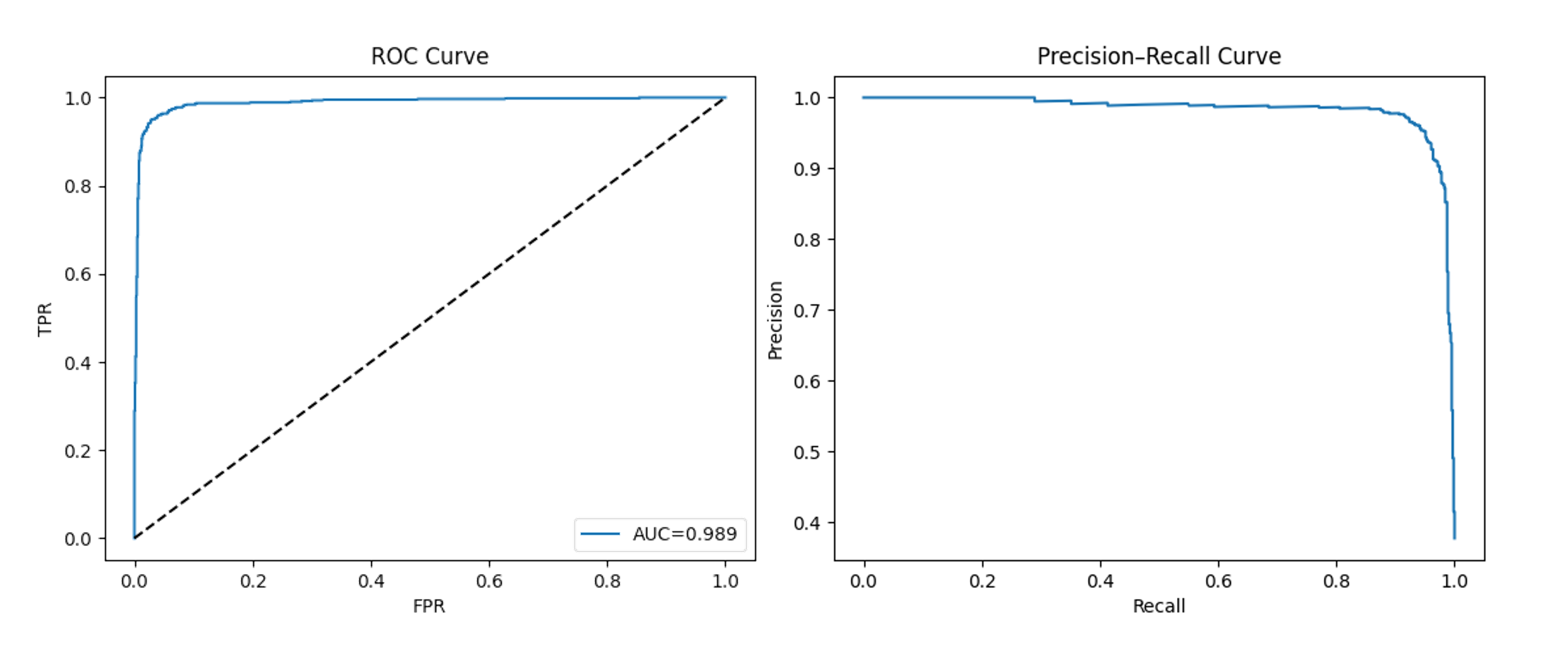}
  \caption{Test-set discrimination. ROC–AUC = 0.989. The PR curve shows sustained high precision across the recall range, which is informative under class imbalance.}
  \label{fig:roc-pr}
\end{figure}

The model achieved ROC–AUC = 0.989 on the test set (Figure~\ref{fig:roc-pr}, left), with the curve closely following the upper-left boundary, indicating very low false-positive rates across thresholds. The PR curve (Figure~\ref{fig:roc-pr}, right) remains near the upper envelope for most of the recall range, demonstrating that high precision is maintained until very high recall. 

\begin{figure}[h!]
    \centering
    \includegraphics[width=1\linewidth]{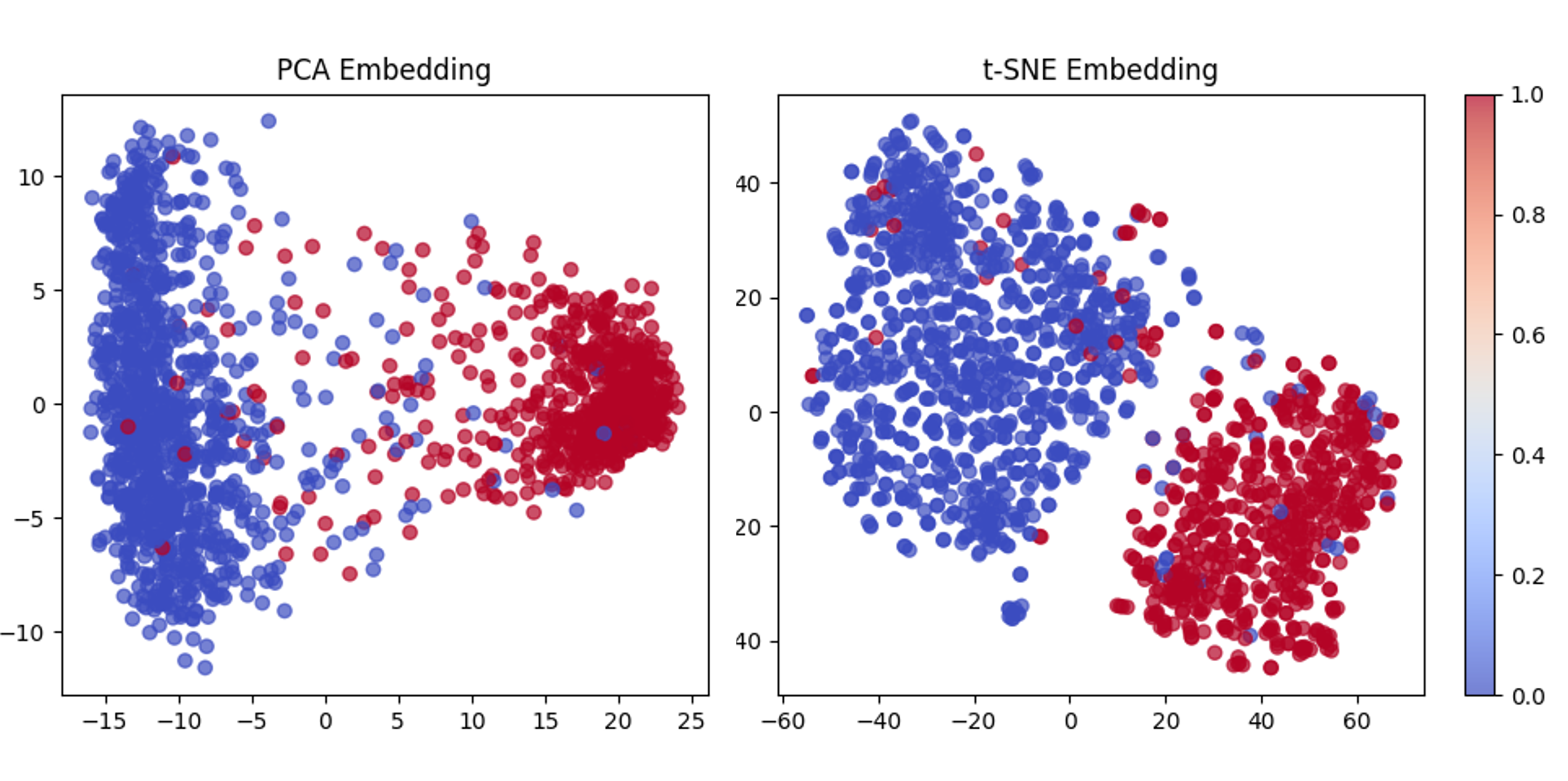}
  \caption{Feature-space visualizations of penultimate-layer embeddings. Left: PCA (2 components). Right: t-SNE (2D, init=random, learning\_rate=auto). Colors denote class labels (red=positive, blue=negative). Both projections show clear class separability.}

  \label{fig:tsne}
\end{figure}

The resulting PCA and t-SNE maps are shown in Figure~\ref{fig:tsne}. PCA already reveals a separation of positive and negative cases, although with partial overlap. t-SNE further accentuates two compact, well-separated clusters, confirming that the learned embeddings discriminate between classes rather than collapsing into undifferentiated space.

\begin{table}[!h]\centering
\caption{Class-wise performance on the test set.}
\label{tab:classwise}
\renewcommand{\arraystretch}{1}
\resizebox{\linewidth}{!}{
    \begin{normalsize}

\begin{tabular}{lccc}
\hline
\hline
Class & Precision & Recall & F1 \\ \hline\hline
Benign & 0.964 & 0.977 & 0.970 \\
Malignant & 0.961 & 0.940 & 0.950  \\ \hline\hline
\end{tabular}
\end{normalsize}}|
\end{table}

\begin{figure}[h!]
    \centering
    \includegraphics[width=1\linewidth]{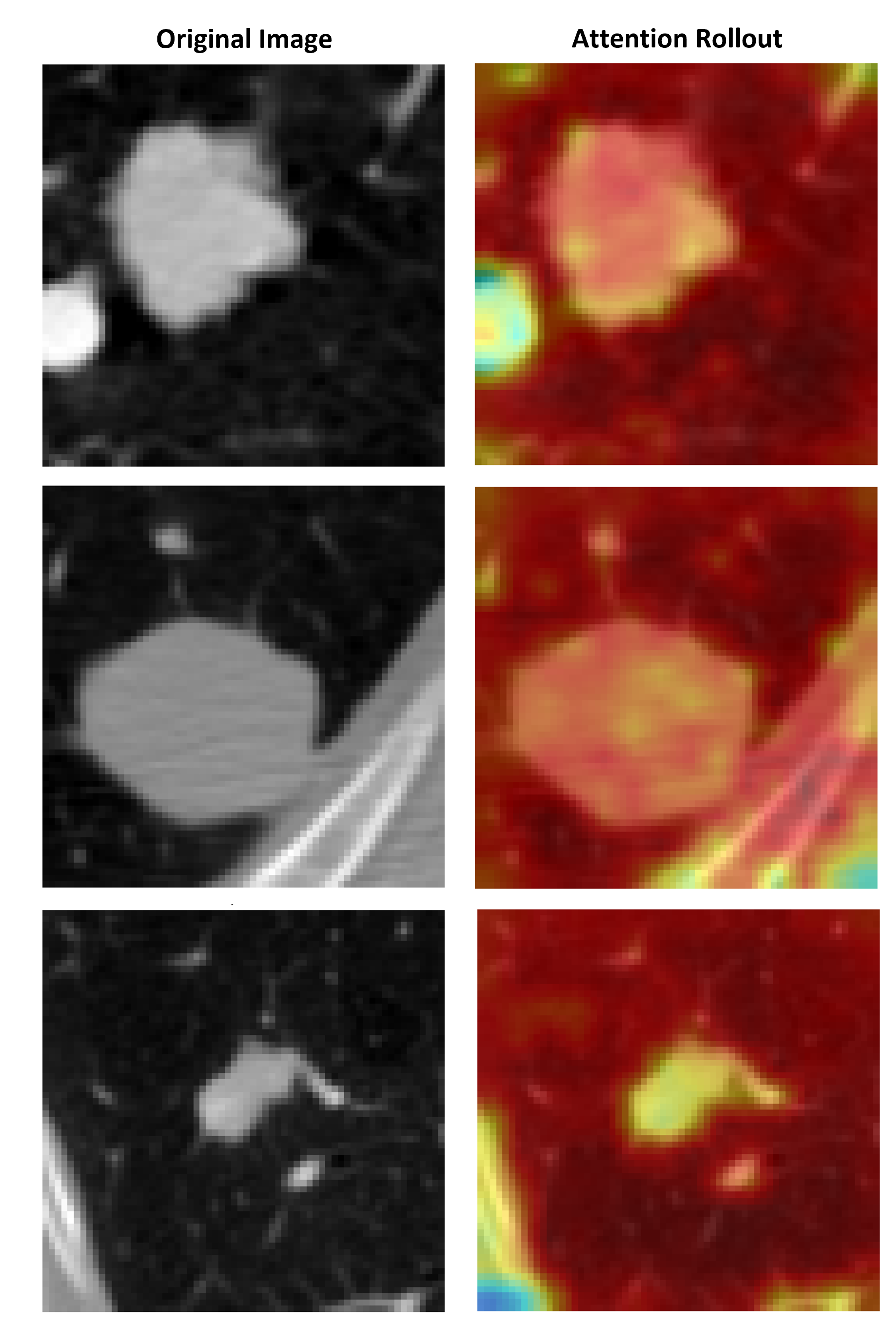}
  \caption{Explainability results for the Vision Transformer (ViT). Left: representative CT images of pulmonary lesions. Right: corresponding attention-rollout heatmaps.}
  \label{fig:Explainability}
\end{figure}

Furthermore, to evaluate model interpretability, we generated attribution maps highlighting the regions most influential in the decision-making process. The attention-rollout heatmaps( see the Figure~\ref{fig:Explainability}) demonstrate that both approaches concentrate on lesion areas and their margins, rather than background textures, indicating that the models attend to clinically meaningful structures. Notably, attention-rollout for the ViT consistently emphasized lesion boundaries, aligning well with radiological markers commonly used in clinical practice. These findings enhance trust in the model predictions and suggest their potential for clinical deployment.

\subsection{Computational Efficiency}
To assess the clinical feasibility of our models, we evaluated inference latency and resource requirements on a moderate target system equipped with a 16 GB GPU, in addition to the high-performance setup used for training. The measured inference times were: Vision Transformer (ViT) $\approx$8–10 ms/sample, YOLOv5s $\approx$3.5–4.7 ms/sample, and 3D Res-UNet $\approx$ 1.6 s/sample (batch size = 1). These results demonstrate that the classification and detection models are capable of near real-time inference, supporting their potential for time-sensitive clinical applications. In contrast, the 3D segmentation model, while accurate, is more computationally demanding, highlighting the trade-off between accuracy and efficiency. 

\section{Limitations and Future Works} \label{mt8}

Although the proposed framework demonstrates promising results, several limitations should be acknowledged.

First, the system requires substantial computational resources, including high-end GPUs and large memory capacity, which may limit deployment in low-resource clinical environments. Optimizing the framework for efficiency and exploring edge or cloud-based implementations will be essential for practical adoption.

Second, the framework has not yet undergone prospective clinical validation. While retrospective evaluation on benchmark datasets (e.g., LUNA16) shows high performance, real-world effectiveness in assisting radiologists remains untested. Future work will include carefully designed clinical studies, involving multi-center trials to evaluate diagnostic accuracy, decision support reliability, and integration into existing workflows.
Third, a formal ablation study is absent. Module-wise evaluation of segmentation, detection, and classification components is required to quantify their independent contributions relative to the full hybrid pipeline. Future studies will perform ablation experiments and compare hybrid versus standalone architectures to understand the performance gains at each stage.

Dataset diversity also remains a limitation. The study primarily relied on the LUNA16 dataset, which lacks broad demographic and scanner variability. Future efforts will focus on testing the framework across larger and more heterogeneous datasets, including NLST, MosMed, and in-house multi-institutional collections, to ensure generalizability.
Finally, deployment considerations such as integration with Picture Archiving and Communication Systems (PACS), automated preprocessing pipelines, and real-time inference constraints have not yet been addressed. Future work will explore these aspects to improve clinical readiness, usability, and scalability.

Together, these directions aim to enhance the robustness, generalizability, and practical applicability of the proposed system for real-world lung cancer diagnosis.

\section{Conclusions} \label{mt10}
In this study, we propose an end-to-end automatic deep neural framework for classifying benign and malignant lung nodules, enabling the detection of lung cancer from CT scans. The ability to detect and classify lung nodules accurately is crucial for early-stage diagnosis, which significantly improves patient outcomes. Our framework integrates three robust architectures: 3D-Res-U-Net, YOLOv5, and ViT, each contributing to different stages of the process. The 3D-Res-U-Net is responsible for precise lung segmentation, YOLOv5 detects the nodules, and ViT is utilized to classify the pathological condition of the nodules. The experimental results demonstrate that the proposed framework reliably segments the lungs, detects nodules, and classifies their pathological condition accurately. These results indicate that our model performs well in handling real-world, complex data. Notably, our ViT architecture achieved an overall accuracy of 96.29\%, surpassing other state-of-the-art architectures' performance in lung nodule classification. This improvement highlights the effectiveness of the Vision Transformer (ViT) in capturing global context and fine details from CT images. We believe that our system has the potential to be a valuable tool in clinical decision support, offering a robust, automated solution for lung cancer classification. Furthermore, it can significantly aid primary decision-making, particularly in resource-limited regions where access to expert radiologists and advanced diagnostic tools may be scarce. By reducing the need for manual intervention and providing timely and accurate results, this system can help bridge the gap in healthcare accessibility.

\section{Declarations}

\subsection*{Ethical approval}
This article does not contain any studies with human participants or animals performed by the authors.

\subsection*{Consent to participate}
Not applicable.

\subsection*{Consent to publish}
Not applicable.

\subsection*{Funding}
None.

\subsection*{Conflict of interest}
The authors declare that they have no conflict of interest.

\subsection*{Use of Generative AI}
Artificial intelligence tools were used only to improve clarity, grammar, and linguistic expression. These tools were not involved in generating scientific content, data interpretation, or analysis. The authors remain fully responsible for the originality, accuracy, and integrity of all intellectual content.

\bibliographystyle{unsrt} 

\textcolor{black}{\bibliography{main}}
\end{document}